\documentclass[letter]{aa}  

\usepackage{graphicx,natbib}
\usepackage{txfonts}
\begin{document}

   \title{Stability and instability of strange dwarfs}

   \author{Francesco Di Clemente \inst{1,2}
          \and
          Alessandro Drago\inst{1,2}
          \and
          Prasanta Char\inst{3}
          \and
          Giuseppe Pagliara\inst{1,2}
          }

   \institute{Dipartimento di Fisica e Scienze della Terra,  Università di Ferrara, Via Saragat 1, I-44122 Ferrara, Italy
         \and
             INFN Sezione di Ferrara, Via Saragat 1, I-44122 Ferrara, Italy
             \and
             Space Sciences, Technologies and Astrophysics Research (STAR) Institute, Université de Liège, Bât. B5a, 4000 Liège, Belgium
             }

 
  \abstract
   {}
   {More than 20 years ago, the existence of stable white dwarfs with a core of strange quark matter was proposed. More recently, via the study of radial modes, it has been concluded instead that such objects are unstable. We aim to clarify this issue.}
   {We investigated the stability of these objects by looking at their radial oscillations while incorporating boundary conditions at the quark-hadron interface, which correspond to either a rapid or a slow conversion of hadrons into quarks.}
   {Our analysis shows that objects of this type are stable if the star is not strongly perturbed and, therefore, ordinary matter cannot transform into strange quark matter because of the Coulomb barrier separating the two components. On the other hand, ordinary matter can be transformed into strange quark matter if the star undergoes a violent process, as in the preliminary stages of a type Ia supernova, and this causes the system to become unstable and collapse into a strange quark star. In this way, the accretion-induced collapse of strange dwarfs can be facilitated, and kilometre-sized objects with sub-solar masses can be produced.}
   {}


   \maketitle
%

\section{\label{sec:introduction}Introduction}
White dwarfs (WDs) are formed when the progenitor star has a mass below around $9\,M_{\odot}$ \citep{Heger:2002by} and runs out of nuclear fuel at the conclusion of its evolutionary cycle, causing its core to collapse as its outer layers expand. Only when the electrons' degeneracy pressure is sufficient to support the structure does it stop collapsing. Depending on the progenitor mass, the nuclear fusion can lead to the production of different nuclei that correspond to different outcomes of the evolutionary cycle: helium (He) WDs, carbon-oxygen (C-O) WDs, and oxygen-neon-magnesium (O-Ne-Mg) WDs. The maximum mass (Chandrasekhar mass) of WDs is about $1.4\,M_{\odot}$ \citep{Chandrasekhar:1931ih}, depending on their composition, and most of them are C-O WDs. 

In 1995 it was proposed that WDs can harbour a core of absolutely stable strange quark matter (satisfying the Bodmer-Witten hypothesis \citep{Bodmer:1971we,Witten:1984rs}) at their centre and that the presence of this strange core can stabilize objects that otherwise would be unstable \citep{Glendenning:1994sp,Glendenning:1994zb}. These objects, called strange dwarfs (SDs), can have radii, masses and an astrophysical evolution that is different from those of normal WDs \citep{Glendenning:1994sp,Glendenning:1994zb,Alford:2017vca}. It was suggested that they form either by cumulating normal nuclear matter on the surface of a strange quark star (QS) or from WDs collecting nuggets of strange quark matter (strangelets) already present in the Galaxy. \citet{Glendenning:1994sp} discussed the radial stability of SDs showing that they can be stable for nuclear matter envelope densities far exceeding the maximum densities of WDs. 

The question of the stability of SDs was reconsidered in \citet{Alford:2017vca}, who show that the eigenvalue of the fundamental radial mode is negative, indicating that such systems are unstable. \footnote{In \citet{Alford:2017vca}, it was suggested that \citet{Glendenning:1994sp,Glendenning:1994zb} mistook the second-lowest eigenmode for
the lowest one.}
Since the two calculations seemed to be based on the same hypothesis, the problem remained unsettled.

In this Letter, we reconsider the stability of SDs by discussing an aspect that was not investigated in the previous papers, that is the appropriate boundary conditions at the interface between nuclear matter and the quark core. In this analysis, we use the formalism developed in \citet{Pereira:2017rmp} and \citet{DiClemente:2020szl}, who show that different boundary conditions need to be applied depending on the rapidity of the conversion of nuclear matter into quark matter and that those boundary conditions can modify the eigenvalue of the radial oscillations and, therefore, the stability of the star. We also show how the traditional stability criterion based on counting the extrema in the mass-radius (MR) plane \citep{zeldovich:1963,bardeen:1966} remains valid, but that it must be implemented by explicitly specifying if, during the radial oscillation, the quark content is kept constant or not.

Finally, the results of our analysis prove to be relevant when discussing the accretion induced collapse (AIC) of WDs \citep{Canal:1990dz} and the possibility of forming kilometre-sized objects of sub-solar mass, such as SAX J1808.4-3658 \citep{DiSalvo:2018mua}. \footnote{It has recently been proposed that the central compact object associated with HESS J1731-347 also has a small radius and a sub-solar mass \citep{doroshenko2022}. Our analysis suggests that object is also a QS \citep{DiClemente:2022wqp}.}

\section{\label{subsec:stability}Structure of strange dwarfs}

The crucial idea that allows SDs to form is that nuclear matter is separated from the core of quark matter by a Coulomb barrier as long as the maximum density of nuclear matter $\varepsilon_\mathrm{t}$ is smaller than the neutron drip density $\varepsilon_\mathrm{drip} \sim 4 \times 10^{11} \mathrm{\,g/cm^{3}}$. Above $\varepsilon_\mathrm{drip}$ free neutrons are present and, since they are not influenced by the Coulomb barrier, they fall into the quark matter core, where they are rapidly absorbed and their quarks deconfined.

\begin{figure}[t]
\begin{centering}
\includegraphics[width=8.6cm]{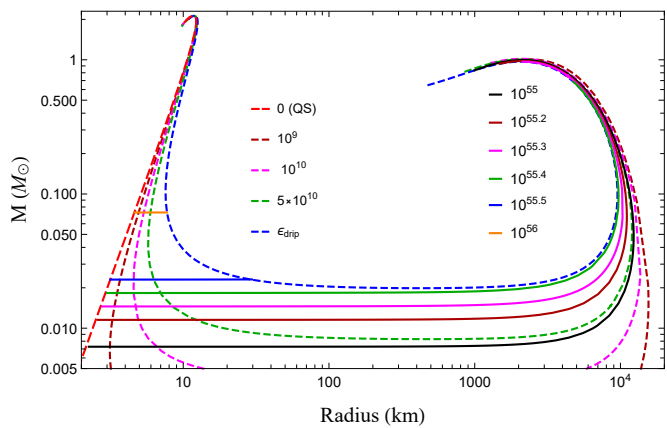}
    \caption{MR sequences. Dashed lines show configurations in which $P_t$ is constant. By increasing $P_0$ (and therefore also $B_\mathrm{core}$) the curves are followed clockwise. The legend indicates $\varepsilon_t$ values in $\mathrm{g/cm^3}$. Solid lines show configurations in which  $B_\mathrm{core}$ is constant. Here, by increasing $P_0$ (and therefore also $P_t$) the curves are followed anti-clockwise. The legend indicates the value of $B_\mathrm{core}$.}
    \label{MRdiagram}
    \end{centering}
    
\end{figure}

The equation of state (EoS) describing this situation and used in \citet{Glendenning:1994sp,Glendenning:1994zb} reads 

\begin{equation}
    \varepsilon(P)=\begin{cases}
    \varepsilon_\mathrm{BPS}(P) & P \leq P_\mathrm{t}\\
    \varepsilon_\mathrm{quark}(P) & P>P_\mathrm{t}
    \end{cases}\label{eq:piecewiseEOS}
\end{equation}
where $\varepsilon_\mathrm{BPS}$ is the Baym-Pethick-Sutherland (BPS) EoS \citep{Baym:1971pw} and $\varepsilon_\mathrm{quark}$ is an EoS for the strange quark matter (e.g. one based on the MIT bag model
\footnote{
The thermodynamic potential of the bag model reads $\Omega (\mu)= -\frac{3}{4 \pi^2} a_4 \, \mu^4 +\frac{3}{4\pi^2}(m_s^2 -4 \Delta_0^2)\mu^2+B$. In our calculations, we used $a_4=1$, gap parameter $\Delta_0 \, = \, 10$ MeV, strange quark mass $m_s\,=\,120$ MeV, and bag constant $B\,=\,135^4 \, $MeV$^4$.}
).  The transition pressure $P_\mathrm{t}\equiv P(R_\mathrm{core})=P(\varepsilon=\varepsilon_\mathrm{t})$ is the pressure at the radius of the interface separating quarks and nuclear matter. We note that the BPS EoS provides
a Chandrasekhar mass of about $1 M_\odot$.

Any value of $\varepsilon_\mathrm{t}$ can be used as long as $\varepsilon_\mathrm{t}<\varepsilon_\mathrm{drip}$. This implies that to define a specific SD stellar configuration, obtained by solving the Tolman–Oppenheimer–Volkoff (TOV) equation \citep{Oppenheimer:1939ne}, one has to define two parameters: $P_\mathrm{t}$ and the central pressure of the star $P_0 \equiv P(r=0)=P(\varepsilon_0)$. This difference with respect to the case for normal WDs plays an important role in the discussion of the stability.

Since the solutions of the TOV equation for SDs are bi-parametric, one has to consider whether the choice of the pair of parameters $(P_0,P_t)$ is the most appropriate. As discussed in a series of papers by \citet{Vartanyan:2009zza,Vartanyan:2012zz}, as long as nuclear matter cannot transform into quark matter, one can define sequences of configurations with the same quark baryon number $B_{\mathrm{ core}}$ and, therefore, choose the two parameters to be $(P_0,B_{\mathrm{ core}})$. 
The quark baryon number reads:
\begin{equation}
   B_\mathrm{core}(\varepsilon_0,\varepsilon_\mathrm{t}) = \int_{0}^{R_\mathrm{core}} {4 \pi r^2 \frac{\rho(r)}{\sqrt{1-2 m(r)/r}}\,dr}\, \label{bcore},
\end{equation}
where $\rho$ is the baryon density. It should be noted that the two choices are not equivalent because if one keeps $P_t$ constant, then $B_{\mathrm{ core}}$ changes while varying $P_0$, describing a situation in which hadrons can deconfine into quarks. If instead $B_{\mathrm{ core}}$ is kept constant, then $P_t$ must increase with increasing $P_0$, describing a situation in which hadrons cumulate on the surface of the strange core without being absorbed. 
\begin{figure}[t]
\begin{centering}
    \includegraphics[width=8.6cm]{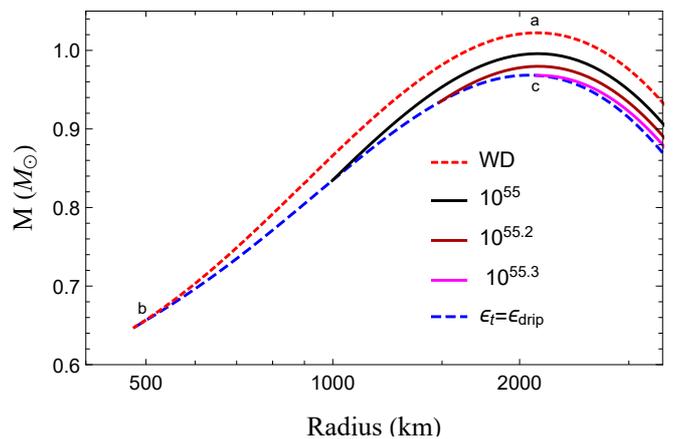}
    \caption{Magnification of the MR sequence close to the Chandrasekhar limit. The notation the same as in Fig.\ref{MRdiagram}. The WD configuration (not shown in Fig.\ref{MRdiagram}) is also displayed.}
    \label{MRzoom}
    \end{centering}
    
\end{figure}

While it is straightforward to build sequences of solutions of the TOV equation obtained by keeping $P_t$ constant and varying $P_0$, it is numerically more cumbersome to produce sequences of solutions specified by a given value of $B_{\mathrm{ core}}$. The reason is that Eq.(\ref{bcore}) provides $B_{\mathrm{ core}}$ as a function of $(P_0,P_t)$, but the inverse relation, providing $P_t$ as a function of $(P_0,B_{\mathrm{ core}})$, can only be obtained numerically \footnote{To this purpose, we found the following expansion useful:
    $\varepsilon_0(B_\mathrm{core},\varepsilon_\mathrm{t})\simeq\varepsilon_\mathrm{t}^\mathrm{Q}/[1-k_1\,\varepsilon_\mathrm{t}^\mathrm{Q}\,B_\mathrm{core}^{2/3}-k_2\,(\varepsilon_\mathrm{t}^\mathrm{Q})^2\,B_\mathrm{core}^{4/3}]\,,$
where $\varepsilon_\mathrm{t}^\mathrm{Q}=\varepsilon_\mathrm{quark}(P_t(\varepsilon_\mathrm{t}))$ and $k_1$, $k_2$ are parameters determined by numerically inverting Eq.(\ref{bcore}). Notice that as long as the quark core is small, $\varepsilon_0\sim \varepsilon_\mathrm{t}^\mathrm{Q}$, but the difference between these two energy densities is crucial for obtaining the correct solutions.}.

In Fig.\ref{MRdiagram} we show the MR relations obtained by keeping either $P_t$ or $B_\mathrm{core}$ fixed. If $B_\mathrm{core}=\mathrm{constant}$, the configurations with smaller $P_0$ values are at the left of the diagram (QSs without any nuclear matter mantle, i.e. $\varepsilon_\mathrm{t}=0$). These pure QS configurations have no unstable mode. With increasing $P_0$, we move anti-clockwise and, according to the general stability criterion \citep{zeldovich:1963,bardeen:1966}, the configurations remain stable until they reach the maximum mass at which the fundamental mode becomes unstable (which happens for $B_\mathrm{core}\lesssim 10^{55.3}$), or until they reach $\varepsilon_\mathrm{t}=\varepsilon_\mathrm{drip}$ (if $B_\mathrm{core}\gtrsim 10^{55.3}$), as discussed in \citet{Glendenning:1994zb,Glendenning:1994sp} and \citet{Vartanyan:2009zza,Vartanyan:2012zz}. Above that value of $\varepsilon_\mathrm{t}$ free neutrons are produced, and the configuration is no longer stable. 

Instead, if $P_t$ is kept constant, we start from the top-right purely nucleonic WD configurations in which $B_\mathrm{core}=0$ (completely stable WD configurations to the right of point (a) in Fig.~\ref{MRzoom}). By increasing $P_0$, we reach point (a) where the fundamental mode becomes unstable. By further increasing $P_0$, we reach $P_t$ at which quarks start being present, and the curve displays a turning point (point (b) in Fig.~\ref{MRzoom} if $\varepsilon_\mathrm{t}=\varepsilon_\mathrm{drip}$). At point (b) the first excited mode also becomes unstable (the path until (b) is followed anti-clockwise). At (c) (this extreme point is reached moving clockwise) the first excited mode again becomes stable while the fundamental mode remains unstable until the minimum mass is reached at a radius of a few hundred kilometres, 
as discussed in \citet{Alford:2017vca}. In conclusion, in both cases the general stability criterion of \citet{zeldovich:1963} and \citet{bardeen:1966} is satisfied.

\begin{figure}[t]
\begin{centering}
   \includegraphics[width=0.5\textwidth]{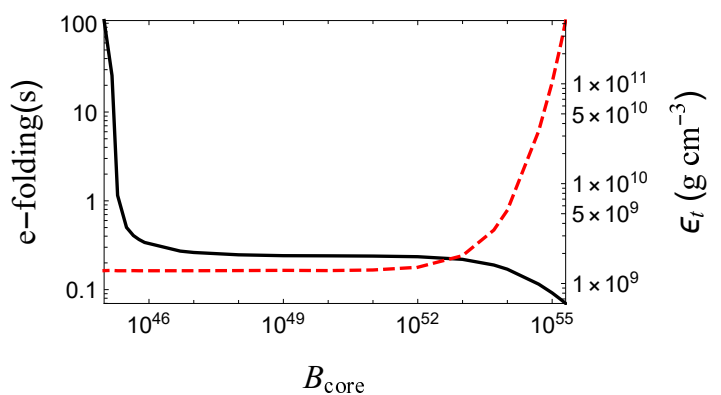}
    \caption{Properties of maximum mass stars as a function of their quark content. The solid black line shows the timescale of the mechanical instability as a function of $B_\mathrm{core}$. The dashed red line shows the transition density.}
    \label{collapse}
    \end{centering}
    
\end{figure}

In Fig.\ref{MRzoom} we show that, as suggested in \citet{Glendenning:1994sp,Glendenning:1994zb}, there are indeed stable configurations of SDs in which the largest density of nuclear matter exceeds that reached in normal WDs. Nevertheless, we note that these configurations are reached only if $B_\mathrm{core}\gtrsim 10^{52}$  (see Fig.~\ref{collapse}), otherwise the structure of an SD with $M\sim M_\odot$ is very similar to that of a WD. Another important difference between the structure of SDs and WDs concerns objects with $M\lesssim 0.1 M_\odot$: their radius can be significantly smaller than that of a WD; the existence of this type of object has been recently suggested in \citet{Kurban:2020xtb}.

\section{\label{subsec:radial}Radial oscillations}

\begin{figure}[t]
\begin{centering}
    \includegraphics[width=8.6cm]{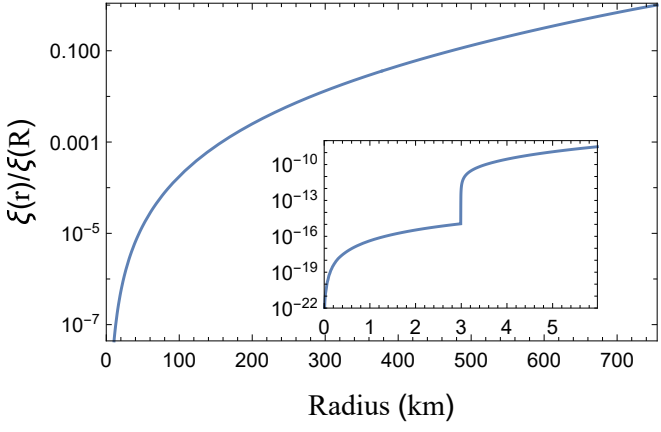}
    \caption{Fundamental eigenfunction of radial modes in the slow scenario in which hadrons do not deconfine into quarks during the oscillation timescale. The star considered here and in Fig. \ref{fast} has M$\,\simeq 0.02 \,\mathrm{M}_\odot$,  $B_\mathrm{core} \simeq 2.69 \times 10^{55}$ and $\varepsilon _t = \varepsilon_\mathrm{drip}$ and is located to the right of the minimum of the dashed blue curve in Fig.~\ref{MRdiagram}. Here, the mode is stable: $\omega^2 = 0.788275 \mathrm{\,Hz^2}$. In the inset plot, the region around $r=r_t$ is magnified: there the eigenfunction has a kink.}
    \label{slow}
    \end{centering}
    
\end{figure}

To check the stability of a configuration, we studied radial oscillations. 
Schwarzschild's line element for a non-rotating symmetric star reads
\begin{equation}
ds^2 = e^{2\phi} dt^2 - e^{2\lambda}dr^2 - r^2(d\theta^2 +\sin^2\theta d\phi^2)\,,
\end{equation}
where $\phi\equiv \phi(r)$ and  $\lambda \equiv \lambda(r)$ are the metric potentials. Using this metric, the differential equation for radial oscillations reads
\begin{equation}
(H \xi')'=-(\omega^2 W + Q)\xi\,,
\label{eq:radial}
\end{equation}
where $\xi(r)$ is the Lagrangian displacement multiplied by $r^2 e^{-\phi}$ and $\omega$ is the characteristic frequency of the mode. Here, 
\begin{align}
H &=  r^{-2} (\varepsilon + P)  e^{\lambda + 3 \phi} c_s^2 \nonumber \\
Q &=  r^{-2} (\varepsilon + P) e^{\lambda + 3 \phi} (\phi'^2 + 4 r^{-1}\phi'-8 \pi e^{2 \lambda}P) \nonumber  \\
W &= r^{-2} (\varepsilon + P)  e^{3\lambda +  \phi}\,,
\label{eq:coefficinets_SL}
\end{align}
where $c_s^2$ is the sound velocity.
The important point is that when 
multiple layers or phase transitions are present, one has to specify boundary conditions on the separating surfaces, as discussed in \citet{Pereira:2017rmp} and \citet{DiClemente:2020szl}. In particular, one has to clarify if, during the timescale of the oscillation, the two components of the fluid can transform one into the other. We are now going to discuss these two possibilities.

\begin{figure}[t]
\begin{centering}
\includegraphics[width=8.6cm]{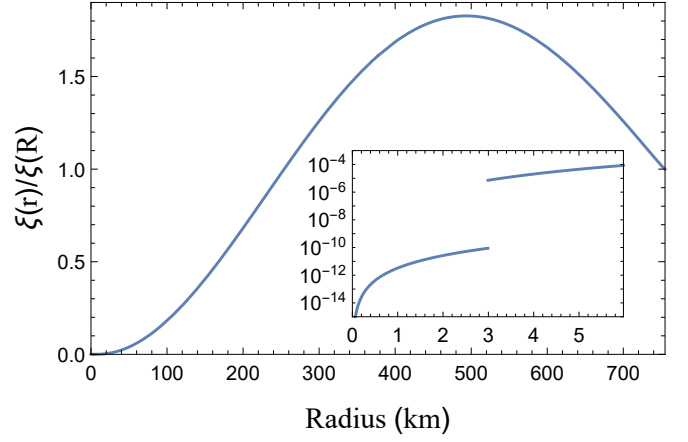}
    \caption{Fundamental eigenfunction in the case of rapid transitions, similar to the scenario discussed in \cite{Alford:2017vca}. Here, the mode is unstable, $\omega^2 = -1.62785 \mathrm{\,Hz^2}$, and the eigenfunction is discontinuous.}
    \label{fast}
    \end{centering}
    
\end{figure}

\paragraph{\label{subsec:slow}Slow transition.}
The slow phase transition scenario occurs when the characteristic timescale of the conversion of one phase into the other is much longer than that of the perturbation. The two phases do not mix, and the volume element close to the surface separating the two phases moves with the interface, expanding and contracting. This case applies to SDs in which $\varepsilon_\mathrm{t} < \varepsilon_\mathrm{drip}$, and it corresponds in the MR diagram to sequences of configurations in which $B_\mathrm{core}$ is kept constant.

The interface conditions for the slow conversion are the continuity of the radial displacement at $r_\mathrm{t}$,
\begin{equation}\label{eq:slow1}
    \left[\xi \right]^+_- \equiv \xi(r_\mathrm{t}^+ )-\xi(r_\mathrm{t}^-)=0\,,
\end{equation}
and the continuity of the Lagrangian perturbation of the pressure,
\begin{equation}\label{eq:slow2}
    \left[\Delta P\right]^+_- = \left[ -e^\phi\, r^{-2}\, \gamma (r) \, P \, \frac{\partial \xi}{\partial r} \right]^+_-=0 \,,
\end{equation}
where $\gamma(r)=(\partial P /\partial \varepsilon)(\varepsilon+P)P^{-1}$ is the relativistic adiabatic index.
By solving Eq.~(\ref{eq:radial}) with conditions (\ref{eq:slow1}) and (\ref{eq:slow2}), we obtain that $\omega^2>0$, and it vanishes at the maximum mass in the MR plane along the curve obtained by keeping $B_\mathrm{core}$ constant, as suggested by the \citet{zeldovich:1963} and \citet{bardeen:1966} criterion. The eigenfunctions are continuous and have a kink at $r_\mathrm{t}$ (see Fig.~\ref{slow}) and the same behaviour is displayed by $\Delta P (r)$.

\paragraph{\label{subsec:fast}Rapid transition.}
When the timescale of the conversion of the two phases is shorter than that of the perturbation, mass transfer between the two phases is possible. The surface separating the two phases is in thermodynamic equilibrium since the conversion rates are very rapid; therefore, Eq. \ref{eq:slow2} also applies to this case.
The only difference with the slow transition case is that the interface condition in Eq. \ref{eq:slow1} becomes
\begin{equation}\label{eq:rapid}
    \left[\xi  + \frac{\gamma P \xi'}{P'}\right]^+_- =0\,,
\end{equation}
such that the eigenfunction has a discontinuity at the interface.

\begin{figure}[t]
\begin{centering}
\includegraphics[width=8.6cm]{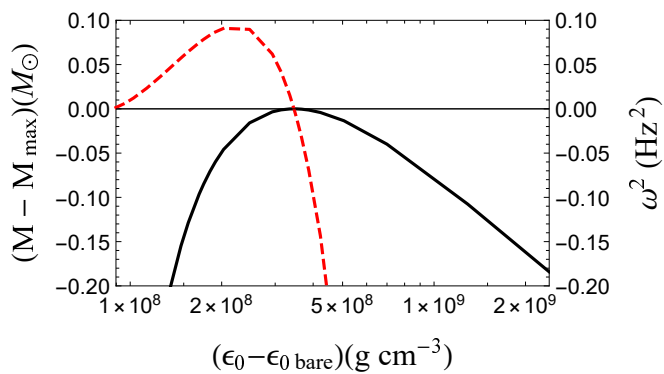}
    \caption{Eigenvalues of the fundamental mode in the slow case (dashed) and masses of SDs having $B_{\text{core}} = 10^{55}$, close to the maximum, $M_{\mathrm {max}}\sim 0.996 M_\odot$ (solid), plotted as functions of the central energy density, $\epsilon_0$. The zero of $\omega^2$ coincides with the maximum mass, and it turns negative at higher densities. Since $\epsilon_0$ remains almost constant in the range displayed in the figure, we show its tiny change with respect to the central density, $\epsilon_{0\,\, \mathrm{bare}}$, of a pure QS that has the same $B_{\text{core}}$.}
    \label{omegafast}
    \end{centering}
    
\end{figure}

The origin of the apparent discrepancy between the results of \cite{Glendenning:1994sp,Glendenning:1994zb} and those of \cite{Alford:2017vca} is now clear. In \cite{Alford:2017vca}, the used EoS is similar to the one discussed in Eq.(\ref{eq:piecewiseEOS}), but a smoothing is introduced \footnote{
The smoothed EoS used in \cite{Alford:2017vca} reads 
$\varepsilon(P)= \left[1-\mathrm{tanh}\left( (P-P_\mathrm{crit})/\delta P\right)\varepsilon_\mathrm{BPS}(P) \right]/2 +
     \left[1+\mathrm{tanh}\left( (P-P_\mathrm{crit})/\delta P\right)\varepsilon_\mathrm{quark}(P) \right]/2$
where $\delta P$ is the transition width.} 
so that there is no sharp discontinuity between the two phases and, crucially, they can transform instantaneously from one into the other. This is similar to the rapid case discussed in this Letter: while in \cite{Alford:2017vca} the eigenfunction does not display a discontinuity at the interface, a very fast increase in its value takes place, the size of which is totally equivalent to that displayed in our Fig.\ref{fast}. On the other hand, no discussion on the boundary conditions at the interface is presented in \cite{Glendenning:1994sp,Glendenning:1994zb}, but most likely the eigenfunction was assumed to be continuous, which is the situation described in our slow scenario. 

The distinction between slow and rapid processes was already introduced in the 1960s (see e.g. \citet{Thorne}), when it was noted that the agreement between the stability analysis based on the solutions of the TOV equation (static analysis) and that based on the study of the radial oscillation equation (dynamic analysis) depends on the use in the eigenvalue equation of an adiabatic index that is derived from the EoS adopted in the static analysis. It is easy to satisfy this agreement in the rapid case. It is, in general, far from easy in the slow case because the
slow adiabatic index is computed by taking imbalances produced by the perturbation into account \citep{Lindblom:2001hd,Drago:2003wg}, while an imbalanced EoS is generally not introduced. Our case is particularly straightforward because the conversion between
hadrons and quarks can take place only on a bi-dimensional surface (and not over an extended volume). It is, therefore, easy to modify both the adiabatic index (this corresponds to modifying the interface conditions; \citet{Pereira:2017rmp}) and the EoS (the slow case corresponds to keeping the quark content completely frozen; \citet{Vartanyan:2009zza,Vartanyan:2012zz}). In this way, we also have a correspondence between static and dynamic analyses in the slow case, as shown in Fig.~\ref{omegafast} (in the rapid case this was already proven by \citet{Alford:2017vca}).

\section{\label{sec:conclusions}Implications for SD collapse}
 The presence of a strange quark matter core in SDs can play a crucial role when the object is strongly perturbed, as occurs in the preliminary stages of a type Ia supernova. In particular, it can help the object collapse instead of following the path that leads to a deflagration. The difficulty in producing an AIC  in a WD is due to the fact that strong nuclear reactions start taking place when the star is close to the Chandrasekhar limit, that is in a situation in which the star is still marginally mechanically stable ($\omega^2 \simeq 0$), and they disrupt the star before AIC takes place \citep{Canal:1990dz}. We have shown in this Letter that the mechanical stability in the case of SDs depends on the possibility of hadrons rapidly converting into quarks. As long as $\varepsilon_t\ll\varepsilon_\mathrm{drip}$ the object is mechanically stable, but if a fluctuation produces matter at densities above $\varepsilon_\mathrm{drip}$, the system becomes unstable. We can estimate this instability by computing the fundamental eigenvalue of a star at the Chandrasekhar limit along a line where $B_\mathrm{core}=$ is constant. In the case of a slow transition $\omega^2=0$; however, in the case of a rapid transition $\omega^2$ is large and negative. In Fig.\ref{collapse} we show the e-folding time = $2 \pi/ |\omega |$; we can see here that for $B_\mathrm{core}\gtrsim 10^{46}$ the typical time of growth of the instability drops well below $1s$, suggesting that the collapse can be faster than the development of the deflagration \citep{Gamezo:2002nc} \footnote{We also computed the e-folding time for a different set of parameters for the quark EoS, i.e. those in \cite{Bombaci:2020vgw}. The results are unchanged.}. In the same figure we also show $\varepsilon_t$, the maximum density reached by the nuclear matter component. As anticipated earlier, the static structure of an SD with $M\sim M_\odot$ does not change until $B_\mathrm{core}\gtrsim 10^{52}$. For smaller values of $B_\mathrm{core}$, $\varepsilon_t$ equals the central density of a WD, indicating that the quark core affects the stability of the star for values of $B_\mathrm{core}$ smaller than those needed to affect its static properties.

A very important question concerning SDs is how they can collect the strange quark matter sitting at their centre. The most obvious answer is that WDs accumulate strangelets during their lifetime. To this purpose, the formula of \citet{Madsen:1988zgf} is very useful: it indicates the rate of strangelets hitting the surface of a star of mass $M$ and radius $R$. If we assume that dark matter is made of strangelets \footnote{The possibility that dark matter is made of strangelets has been discussed in a few recent papers \citep{Burdin:2014xma,Jacobs:2014yca,Caloni:2021bwp} and the search for these objects is at the core of many observation campaigns \citep{Bacholle:2020emk,POEMMA:2020ykm,JEM-EUSOColaboration:2014pci}.} with a density equal to that of dark matter in the galactic halo,  $\rho_\mathrm{DM}\sim 10^{-24}\,\mathrm{g/cm^3}$, and a velocity of $250\, \mathrm{km/s}$, the rate reads
\begin{equation}
F\sim (1.39 \times 10^{30} \mathrm{s}^{-1}) A^{-1}(M/M_\odot)(R/R_\odot)\, .
\end{equation}
A WD close to the Chandrasekhar mass with a radius of $\sim 3000$ km can in 5 Gy reach $B_\mathrm{core} \sim 10^{45}$, just slightly smaller than the size needed to affect the dynamics of AIC. We can nonetheless see that the density distribution of dark matter grows rapidly towards the centre (see for example \citet{Navarro:1996gj}), meaning that in the most central regions of the Milky Way $B_\mathrm{core}$ can easily be large enough to favour AIC. This can be useful to justify schemes in which it is assumed that the AIC is very common close to the galactic centre, such as in the scenario developed to interpret the gamma-ray excess signal from the galactic centre \citep{Gautam:2021wqn}. 

It is interesting to compare the scenario we are describing with the one proposed in \cite{Leung:2013pra}, where the impact of dark matter on the AIC of WDs is also discussed. The main difference is that in \cite{Leung:2013pra} dark matter and normal matter cannot transform into each other (which is the cause of the instability described in this Letter) and, therefore, a huge amount of dark matter is needed to affect the structure of the WD only through gravity. This is consistent with our Fig.\ref{collapse}, where we show that a core of strange quark matter can destabilize the star (by mixing with ordinary matter) even if its size is much smaller than that needed to influence the structure of the star through gravity. In this way, AIC is facilitated and requires an amount of strange quark matter totally compatible with the density of dark matter in our galaxy, as discussed above.

Another interesting outcome of the AIC of SDs is the possibility of producing kilometre-sized objects with sub-solar masses. We note that the outcome of an AIC of an SD is a QS and not a neutron star because the large core of strange quark matter would rapidly transform nucleons into quarks \citep{Drago:2015fpa}. In order to estimate the mass of the object produced after the collapse, we must first note that in our calculation we have used the BPS EoS for simplicity, but a more realistic calculation based on a C-O WD would give a Chandrasekhar mass of $\sim 1.4 M_\odot$. In the case of an AIC of a WD into a neutron star, the extra binding in the more compact object corresponds to roughly $0.1 M_\odot$, but if a QS forms from an SD, its gravitational mass can be lower than $\sim 1.1 M_\odot$ \citep{Bombaci:2020vgw}. Moreover, the huge amount of energy released in the conversion can increase the amount of mass ejected during the AIC \citep{Sharon:2019mji}, so the final configuration can be a QS with a radius $R\lesssim 11 \mathrm{km}$ and a mass lower than that of the Sun. It is worth noting that SAX J1808.4-3658 is reported to have a small radius \citep{Li:1999wt,Poutanen:2003yd,Leahy:2007fb}, an accelerated cooling \citep{Heinke:2008vj}, and a sub-solar mass \citep{DiSalvo:2018mua}, all features suggesting that it is a QS, possibly produced through an AIC of an SD.
\\
\begin{acknowledgements} 
We thank Micha{\l}  Bejger for several useful discussions during the preparation of this work.
PC is currently supported by the Fonds de la Recherche Scientifique-FNRS, Belgium, under grant No. 4.4503.19.

\end{acknowledgements}

\bibliographystyle{aa} 
\bibliography{biblio}

\begin{thebibliography}{41}
\expandafter\ifx\csname natexlab\endcsname\relax\def\natexlab#1{#1}\fi

\bibitem[{Alford {et~al.}(2017)Alford, Harris, \& Sachdeva}]{Alford:2017vca}
Alford, M.~G., Harris, S.~P., \& Sachdeva, P.~S. 2017, Astrophys. J., 847, 109

\bibitem[{Bacholle {et~al.}(2021)}]{Bacholle:2020emk}
Bacholle, S. {et~al.} 2021, Astrophys. J. Suppl., 253, 36

\bibitem[{{Bardeen} {et~al.}(1966){Bardeen}, {Thorne}, \&
  {Meltzer}}]{bardeen:1966}
{Bardeen}, J.~M., {Thorne}, K.~S., \& {Meltzer}, D.~W. 1966, \apj, 145, 505

\bibitem[{Baym {et~al.}(1971)Baym, Pethick, \& Sutherland}]{Baym:1971pw}
Baym, G., Pethick, C., \& Sutherland, P. 1971, Astrophys. J., 170, 299

\bibitem[{Bodmer(1971)}]{Bodmer:1971we}
Bodmer, A.~R. 1971, Phys. Rev. D, 4, 1601

\bibitem[{Bombaci {et~al.}(2021)Bombaci, Drago, Logoteta, Pagliara, \&
  Vida\~na}]{Bombaci:2020vgw}
Bombaci, I., Drago, A., Logoteta, D., Pagliara, G., \& Vida\~na, I. 2021, Phys.
  Rev. Lett., 126, 162702

\bibitem[{Burdin {et~al.}(2015)Burdin, Fairbairn, Mermod, Milstead, Pinfold,
  Sloan, \& Taylor}]{Burdin:2014xma}
Burdin, S., Fairbairn, M., Mermod, P., {et~al.} 2015, Phys. Rept., 582, 1

\bibitem[{Caloni {et~al.}(2021)Caloni, Gerbino, \& Lattanzi}]{Caloni:2021bwp}
Caloni, L., Gerbino, M., \& Lattanzi, M. 2021, JCAP, 07, 027

\bibitem[{Canal {et~al.}(1990)Canal, Isern, \& Labay}]{Canal:1990dz}
Canal, R., Isern, J., \& Labay, J. 1990, Ann. Rev. Astron. Astrophys., 28, 183

\bibitem[{Casolino {et~al.}(2015)}]{JEM-EUSOColaboration:2014pci}
Casolino, M. {et~al.} 2015, Exper. Astron., 40, 19

\bibitem[{Chandrasekhar(1931)}]{Chandrasekhar:1931ih}
Chandrasekhar, S. 1931, Astrophys. J., 74, 81

\bibitem[{Di~Clemente {et~al.}(2022)Di~Clemente, Drago, \&
  Pagliara}]{DiClemente:2022wqp}
Di~Clemente, F., Drago, A., \& Pagliara, G. 2022 [\eprint[arXiv]{2211.07485}]

\bibitem[{Di~Clemente {et~al.}(2020)Di~Clemente, Mannarelli, \&
  Tonelli}]{DiClemente:2020szl}
Di~Clemente, F., Mannarelli, M., \& Tonelli, F. 2020, Phys. Rev. D, 101, 103003

\bibitem[{Di~Salvo {et~al.}(2019)Di~Salvo, Sanna, Burderi, Papitto, Iaria,
  Gambino, \& Riggio}]{DiSalvo:2018mua}
Di~Salvo, T., Sanna, A., Burderi, L., {et~al.} 2019, Mon. Not. Roy. Astron.
  Soc., 483, 767

\bibitem[{Doroshenko {et~al.}(2022)Doroshenko, Suleimanov, Puehlhofer, \&
  Santangelo}]{doroshenko2022}
Doroshenko, V., Suleimanov, V., Puehlhofer, G., \& Santangelo, A. 2022, Nature
  Astronomy [\eprint{https://doi.org/10.1038/s41550-022-01800-1}]

\bibitem[{Drago {et~al.}(2005)Drago, Lavagno, \& Pagliara}]{Drago:2003wg}
Drago, A., Lavagno, A., \& Pagliara, G. 2005, Phys. Rev. D, 71, 103004

\bibitem[{Drago \& Pagliara(2015)}]{Drago:2015fpa}
Drago, A. \& Pagliara, G. 2015, Phys. Rev. C, 92, 045801

\bibitem[{Gamezo {et~al.}(2003)Gamezo, Khokhlov, Oran, Chtchelkanova, \&
  Rosenberg}]{Gamezo:2002nc}
Gamezo, V.~N., Khokhlov, A.~M., Oran, E.~S., Chtchelkanova, A.~Y., \&
  Rosenberg, R.~O. 2003, Science, 299, 77

\bibitem[{Gautam {et~al.}(2022)Gautam, Crocker, Ferrario, Ruiter, Ploeg,
  Gordon, \& Macias}]{Gautam:2021wqn}
Gautam, A., Crocker, R.~M., Ferrario, L., {et~al.} 2022, Nature Astron., 6, 703

\bibitem[{Glendenning {et~al.}(1995{\natexlab{a}})Glendenning, Kettner, \&
  Weber}]{Glendenning:1994sp}
Glendenning, N.~K., Kettner, C., \& Weber, F. 1995{\natexlab{a}}, Phys. Rev.
  Lett., 74, 3519

\bibitem[{Glendenning {et~al.}(1995{\natexlab{b}})Glendenning, Kettner, \&
  Weber}]{Glendenning:1994zb}
Glendenning, N.~K., Kettner, C., \& Weber, F. 1995{\natexlab{b}}, Astrophys.
  J., 450, 253

\bibitem[{Heger {et~al.}(2003)Heger, Fryer, Woosley, Langer, \&
  Hartmann}]{Heger:2002by}
Heger, A., Fryer, C.~L., Woosley, S.~E., Langer, N., \& Hartmann, D.~H. 2003,
  Astrophys. J., 591, 288

\bibitem[{Heinke {et~al.}(2009)Heinke, Jonker, Wijnands, Deloye, \&
  Taam}]{Heinke:2008vj}
Heinke, C.~O., Jonker, P.~G., Wijnands, R., Deloye, C.~J., \& Taam, R.~E. 2009,
  Astrophys. J., 691, 1035

\bibitem[{Jacobs {et~al.}(2015)Jacobs, Starkman, \& Lynn}]{Jacobs:2014yca}
Jacobs, D.~M., Starkman, G.~D., \& Lynn, B.~W. 2015, Mon. Not. Roy. Astron.
  Soc., 450, 3418

\bibitem[{Kurban {et~al.}(2022)Kurban, Huang, Geng, \& Zong}]{Kurban:2020xtb}
Kurban, A., Huang, Y.-F., Geng, J.-J., \& Zong, H.-S. 2022, Phys. Lett. B, 832,
  137204

\bibitem[{Leahy {et~al.}(2008)Leahy, Morsink, \& Cadeau}]{Leahy:2007fb}
Leahy, D.~A., Morsink, S.~M., \& Cadeau, C. 2008, Astrophys. J., 672, 1119

\bibitem[{Leung {et~al.}(2013)Leung, Chu, Lin, \& Wong}]{Leung:2013pra}
Leung, S.~C., Chu, M.~C., Lin, L.~M., \& Wong, K.~W. 2013, Phys. Rev. D, 87,
  123506

\bibitem[{Li {et~al.}(1999)Li, Bombaci, Dey, Dey, \& van~den
  Heuvel}]{Li:1999wt}
Li, X.~D., Bombaci, I., Dey, M., Dey, J., \& van~den Heuvel, E. P.~J. 1999,
  Phys. Rev. Lett., 83, 3776

\bibitem[{Lindblom \& Owen(2002)}]{Lindblom:2001hd}
Lindblom, L. \& Owen, B.~J. 2002, Phys. Rev. D, 65, 063006

\bibitem[{Madsen(1988)}]{Madsen:1988zgf}
Madsen, J. 1988, Phys. Rev. Lett., 61, 2909

\bibitem[{Navarro {et~al.}(1997)Navarro, Frenk, \& White}]{Navarro:1996gj}
Navarro, J.~F., Frenk, C.~S., \& White, S. D.~M. 1997, Astrophys. J., 490, 493

\bibitem[{Olinto {et~al.}(2021)}]{POEMMA:2020ykm}
Olinto, A.~V. {et~al.} 2021, JCAP, 06, 007

\bibitem[{Oppenheimer \& Volkoff(1939)}]{Oppenheimer:1939ne}
Oppenheimer, J.~R. \& Volkoff, G.~M. 1939, Phys. Rev., 55, 374

\bibitem[{Pereira {et~al.}(2018)Pereira, Flores, \& Lugones}]{Pereira:2017rmp}
Pereira, J.~P., Flores, C.~V., \& Lugones, G. 2018, Astrophys. J., 860, 12

\bibitem[{Poutanen \& Gierlinski(2003)}]{Poutanen:2003yd}
Poutanen, J. \& Gierlinski, M. 2003, Mon. Not. Roy. Astron. Soc., 343, 1301

\bibitem[{Sharon \& Kushnir(2020)}]{Sharon:2019mji}
Sharon, A. \& Kushnir, D. 2020, Astrophys. J., 894, 146

\bibitem[{Thorne(1966)}]{Thorne}
Thorne, K.~S. 1966, The General-Relativistic Theory of Stellar Structure and
  Dynamics, in Proceedings of the International School of Physics Enrico Fermi.
  Course XXXV, at Varenna, Italy, July 12-24 1965. Academic Press, New York,
  166

\bibitem[{Vartanyan {et~al.}(2009)Vartanyan, Hajyan, Grigoryan, \&
  Sarkisyan}]{Vartanyan:2009zza}
Vartanyan, Y.~L., Hajyan, G.~S., Grigoryan, A.~K., \& Sarkisyan, T.~R. 2009,
  Astrophysics, 52, 300

\bibitem[{Vartanyan {et~al.}(2012)Vartanyan, Hajyan, Grigoryan, \&
  Sarkisyan}]{Vartanyan:2012zz}
Vartanyan, Y.~L., Hajyan, G.~S., Grigoryan, A.~K., \& Sarkisyan, T.~R. 2012,
  Astrophysics, 55, 98

\bibitem[{Witten(1984)}]{Witten:1984rs}
Witten, E. 1984, Phys. Rev. D, 30, 272

\bibitem[{Zel'dovich(1963)}]{zeldovich:1963}
Zel'dovich, Y.~B. 1963, Voprosy kosmogonii, 9, 36

\end{thebibliography}
\end{document}